\documentclass[aps,a4paper,twocolumn,nofootinbib,showpacs,showkeys,preprintnumbers,amsmath,amssymb,10pt]{revtex4-1}

\usepackage{amsmath,amssymb}

\usepackage{graphicx}
\usepackage[mathscr]{eucal}
\usepackage{color}

\newcommand{\bea}{\begin{eqnarray}}	
\newcommand{\eea}{\end{eqnarray}}

\begin{document}

\title{A simple proof of orientability in colored group field theory} 
\author{Francesco Caravelli}


\affiliation{University of Waterloo,\\ Waterloo, Ontario N2L 3G1, Canada\\ and}
\affiliation{Perimeter Institute for Theoretical Physics,\\ Waterloo, Ontario N2L 2Y5, Canada\\}
\affiliation{Max Planck Institute for Gravitational Physics (Albert Einstein Institute),\\ Am
M\"{u}hlenberg 1, D-14476 Golm, Germany}
\email{fcaravelli@perimeterinstitute.ca}
\date{\small\today}

\begin{abstract}
In this short note we use results from the theory of crystallizations to prove that color in group field theories garantees orientability of the piecewise linear pseudo-manifolds 
associated to each graph generated perturbatively. The origin of orientability is the presence of two interaction vertices.
\end{abstract}

\maketitle


\section{Introduction}
There has been recently a growth of interest in group field theories 
\cite{laurentgft,quantugeom2} and there are many reasons for this to happen. Group field theories (GFT) are a generalization of matrix models to higher dimensions\cite{Brezin:1977sv,mm} 
and a generalization of tensor models as well\cite{tensmodl}. Moreover, GFT are known to generate the partition function of Spin Foams, thus having a direct relation with
Loop Quantum Gravity \cite{lqg}.

It is known that matrix models have a 
topological expansion in which the genus, the only topological 
invariant needed to characterize orientable surfaces, plays the role of the parameter of this expansion. 
Roughly speaking a $n$-dimensional group field theory has a vertex associated to an $n$-simplex and a propagator which glues the $(n-1)$-simplices. 
Feynman diagrams of a $n$-dimensional group field theory can be interpreted as gluings of simplices and then have the interpretation of 
piecewise linear (PL) manifolds.  A \textit{colored} version of group field theory (cGFT) has been 
introduced recently\cite{color,PolyColor,sefu2}. One important reason to introduce color in the diagrams is that it is possible to have a better control
over the perturbatively generated singularities of GFT. The challenge in these models is to obtain a topological expansion as in the 2-dimensional 
case \cite{sefu3,FreiGurOriti,sefu1,smerlak}. Remarkably, it has been shown
\cite{1on} that spheres dominates the partition function in any dimension. In order to achieve this result, techniques from the theory of crystallizations have been used.
In fact, colored $n$-graphs are well known in mathematics as \textit{gems}: graph-encoded manifolds \cite{Pezzana,Lins}. 
In this paper we use the results in this field of mathematics to show that the growth of interest in colored models is not unjustified: colored models generate
orientable pseudomanifolds in any number of dimensions. Many of the theorems we will use were known for long time in the context of crystallization and here 
we report briefly these results. The outcome of this note is that the generation of pseudo-manifolds is due to the color, while the orientability in the colored versions of group
field theory models is due to the presence of two different vertices (clockwise and anti-clockwise). In the following we will focus on the Boulatov model, but the result
is more general because it relies only on the presence of vertices of opposite orientation in the perturbative expansion of the partition function.
The paper is organized as follows: in section II we recall the 
colored Boulatov model and its
standard interpretation. In section III we review basic results in the field of 3-gems and crystallizations. 
We will use some of these results in section IV to prove the orientability of 
simplicial complexes generated perturbatively by the colored Boulatov model. Conclusions follow.
\section{The Colored Boulatov Model}
In this section we introduce the colored Boulatov model\cite{GFT}\cite{color}.
Let us consider a compact Lie group $H$, denote $h$ its elements,
$e$ the unit element, and $\int dh$ the integral with respect to the Haar measure of the group.

In 3 dimensions we introduce two fields, $\bar \psi^i$ and $\psi^i$,  $i=0,1,2,3$ be four couples of complex 
scalar (or Grassmann) fields over three copies of $G$, $\psi^i:G\times G\times G 
\rightarrow \mathbb{C}$. The index $i$ runs from$=0$ to $n+1$, where $n$ is the number of dimensions, and
the $\psi$ and $\bar \psi$ are functions of $n$ copies of the group. In the fermionic version of the theory the
indices $i$ can be seen as the dependence of the field from a (global) gauge group $SU(N)$, where $N=n+1$.
We denote $\delta^\Lambda(h)$ the 
regularized delta function over $G$ with some cutoff $\Lambda$ such that 
$\delta^\Lambda(e)$ is finite, but diverges when $\Lambda$ goes to infinity. 
A feasible regularization is given, for instance for the group $G=SU(2)$, by
\bea
  \delta^\Lambda(h) = \sum_{j=0}^{\Lambda} (2j+1) \chi^{j}(h)  .
\eea
where $\chi^j(h)$ is the character of $h$ in the representation $j$. The path integral for the colored Boulatov model over $G$ is:
\begin{eqnarray}
Z(\lambda,\bar\lambda)&=& e^{-F(\lambda,\bar\lambda)} \nonumber \\
&=& \int \prod_{i=0}^4 d\mu_P(\bar \psi^i,\psi^i) \; e^{-S^{int}(\bar \psi^i,\psi^i)} \; ,
\label{eq:part}
\end{eqnarray}
where the Gaussian measure $P$ is chosen such that:
$$\int \prod_{i=0}^4 d\mu_P(\bar \psi^i,\psi^i)=1\ ,$$
and:
\bea
&& P_{h_{0}h_{1}h_{2} ; h_{0}'h_{1}'h_{2}'} = \nonumber \\
&& = \int d\mu_P(\bar \psi^i,\psi^i) \; \bar\psi^i_{h_{0}h_{1}h_{2}} \psi^i_{h_{0}'h_{1}'h_{2}'} = \nonumber \\
&& = \int dh \;  \delta^{\Lambda}\bigl( h_{0} h (h_{0}')^{-1} \bigr) \delta^{\Lambda}\bigl( h_{1} h (h_{1}')^{-1} \bigr) \delta^{\Lambda}\bigl( h_{2} h (h_{2}')^{-1} \bigr) \; \nonumber  ,
\eea
The fermionic colored model has two types interactions, a ``clockwise'' and an ``anti-clockwise'', and one is obtained from the other one by
conjugation in the internal group color $SU(N)$, where $N$ is 4 in 3 dimensions, one for each face of the 3-simplex
\footnote{It should be mentioned that 
also in the bosonic version\cite{sefu2} there is a clockwise and anti-clockwise interaction. In that case the the types of interactions are motivated by the fact that its
introduction has a nice combinatorial definition of homology\cite{PolyColor}}. 
For convenience we denote $\psi(h,p,q)=\psi_{hpq}$. Invariance under global rotations in the internal
color group require at least two interactions:  
\begin{eqnarray}
& S^{int} =  \frac{\lambda}{\sqrt{\delta^\Lambda (e) } } \int (dh)^6 \psi^0   \psi^1 \psi^2  \psi^3  \nonumber \\
& + \frac{\bar\lambda}{\sqrt{\delta^\Lambda(e)}}  \int (dh)^6 \bar \psi^0  \bar \psi^1  \bar \psi^2  \bar \psi^3 
\label{eq:interaction}
\end{eqnarray}
where we omitted the internal structure of the group elements of the fields $\psi^i$ and $\bar \psi^i$. In order to make the notation clearer 
(already the orientation of the colors is sufficient to distinguish the two vertices), we call ``red'' the vertex involving the $\psi$'s and
``black'' the one involving the $\bar \psi$'s. Thus any line coming out of a cGFT vertex has a color $i$. 

The group elements $h_{ij}$ in eq. (\ref{eq:interaction}) are associated to the  propagators (represented 
as solid lines), and glue two vertices with opposite orientation. The vertex can be seen as the dual of a tetrahedron and 
its lines represent the triangles which form the tetrahedron. 
\begin{figure}[htb]
\begin{center}
\includegraphics[scale=0.5]{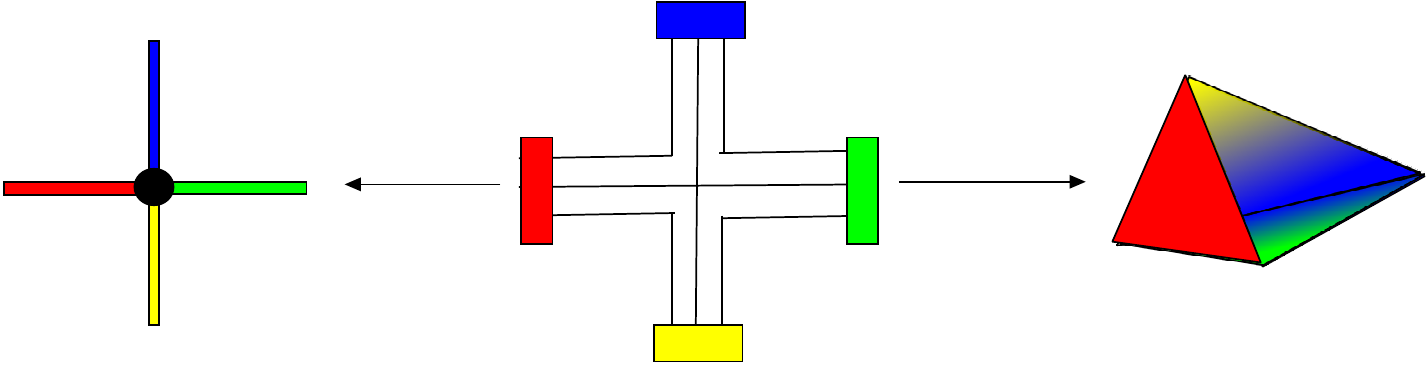}\hspace{0.5cm} 
\includegraphics[scale=0.5]{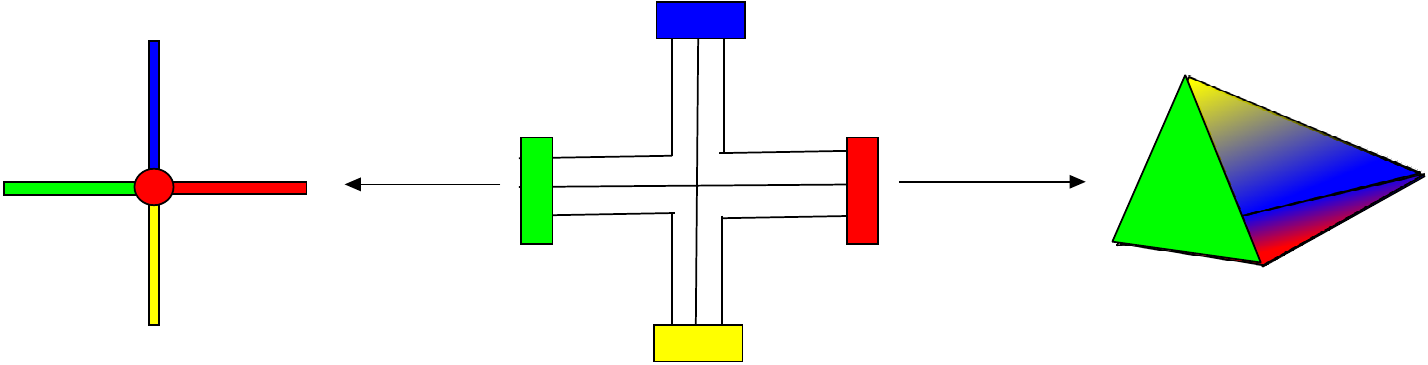}\hspace{0.5cm} 
\caption{Colored GFT red and black vertices.}
\label{fig:vertex}
\end{center}
\end{figure}
\begin{figure}[htb]
\begin{center}
\includegraphics[scale=0.6]{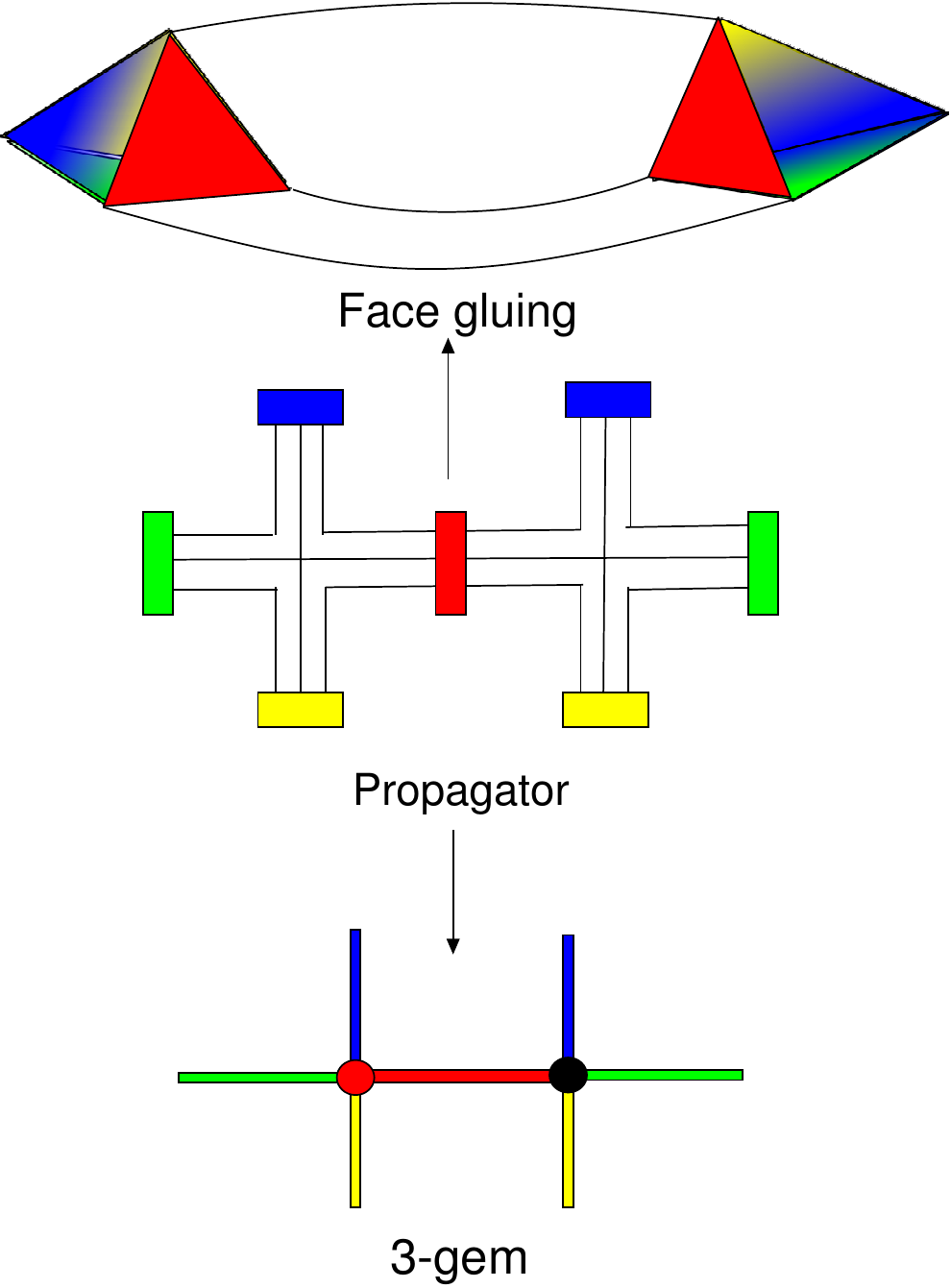}\hspace{0.5cm}
\caption{A gluing using a colored propagator.}
\label{fig:prop}
\end{center}
\end{figure}
Each propagators is decomposed into three {\it parallel} strands 
which are associated to the three arguments of the fields, i.e. the 1-dimensional elements of the 1-skeleton of the tetrahedron which bound every face. These
are associated to the edges of the tetrahedron. 
A colored line represents the gluing of 
two tetrahedra (of opposite orientations) along triangles of the same color as in Fig. (\ref{fig:prop}).

It is easy to understand that a cGFT graph can be seen either as a stranded graph (using the vertex and the propagators as depicted
in Fig. \ref{fig:vertex}) or as a ``colored graph'' with (colored) solid lines, 
and two classes of oriented vertices. In this paper we consider only vacuum graphs, i.e. all the vertices of the graphs are 4-valent and we deal only with connected graphs (thus with the logarithm of the partition
function (\ref{eq:part})).
The lines of a vacuum cGFT graph $\Gamma$ have two natural orientations given by the fact that only vertices of opposite orientations can be glued. It is easy to see that a vacuum cGFT graph must have the same number
of black and red vertices. 
For any graph $\Gamma$, we denote $n$ as the number of vertices, $l$ as the lines of $\Gamma$, and we define as \textit{faces} (not to be confused with the faces of the tetrahedron!), $\mathscr F_{\Gamma}$, as any closed strand in the Feynman graph of a GFT. 
Thus a generic vacuum Feynman amplitude of the theory can be written as:
\bea\label{eq:ampli}
 \mathscr A  = \frac{(\lambda\bar\lambda)^{\frac{n}{2}}} { [\delta^N(e)]^{\frac{n}{2}}} 
\int \prod_{l\in \Gamma} dh_{l} 
\prod_{f\in \mathscr F_{\Gamma}} \delta^\Lambda_{f}(\prod_{l_0 \in f }^{\rightarrow} h_{l_0}^{\sigma(l_0,f)} ) ,
\eea
where $l_0$ is a line associated to a face $f$ and $\sigma(l_0,f)$ is alternatively $+1$ or $-1$ depending on the orientation. In the following we will assume that an orientation is fixed. Because of the properties 
of $\delta's$ the orientation does not affect the amplitude.
To each colored graph associated to an amplitude of the colored Boulatov model it is possible to associate \textit{bubbles} by removing all the edges of one color. We call $\mathscr B_{i_1, \cdots,i_k}$ the set of $k$-bubbles associated to the deletion of $n-k$ colors. 
In 3-dimensions, for instance, 3-bubbles have 3-colors (surfaces), 2-bubbles have 2 colors (lines) and so on and so forth. Bubbles play a special role in the theory, since they discriminate
manifold from pseudo-manifolds (see next section for the same result in the theory of 3-gems). 
\section{A survey of Graph-Embedded Manifolds results}
In this section we review some basic results in the field of 3-gems and make a dictionary between the two literatures, 
as colored group field theory can gain much from the results obtained in all the years of research in such field.\\\ \\
Let $\Gamma$ be a finite, edge-colored graph, parallel edges allowed. A $k$-\textit{residue} of $\Gamma$, $k\in \textbf{N}$ is a connected component of subgraph of 
$\Gamma$ induced by k color classes (this is what in colored group field theory are called \textit{bubbles}). 
These graphs represent a piecewice linear manifold in the following sense (a \textit{pseudo}-complex) \cite{surveyger}. 
A $n$-regular $n$-colored graph is an edge-colored graph which has a nodes of degree $n$.
To a couple $(\Gamma,\gamma)_{n+1}$ there is an associated 
\textit{pseudo}-complex $K(\Gamma)$ given by the following construction. Take an $n$-simplex $\sigma^n$ for each $V(\Gamma)$ and label its vertices $\Delta_n$. If $x$,$y$ in $V(\Gamma)$ are joined by an edge,
then attach the $(n-1)$-faces of their associated simplices. This is the same interpretation given to attaching faces of $n$-simplices in a $n$-dimensional group field theory. We denote $|\Gamma|$ the pseudo-complex associated
with the colored graph $\Gamma$. \\\ \\
\textbf{Lemma 1} For any PL $n$-manifold $\mathscr M$ there exist a (n+1)-graph $\Gamma$ such that $|\Gamma| \backsimeq \mathscr M$.\\\ \\
We now restrict to the case of 3-dimensions and list some of the basic results\cite{Lins}.\\\ \\
Let $\Gamma$ be a 4-edge-colored 4-graph and denote by $v$, $e$, $b$, $t$ respectively the number of vertices (0-\textit{residues}),
edges (1-\textit{residues}), 2-\textit{residues} and 3-\textit{residues}. \\\ \\
\textbf{Definition} A \textit{3-gem} (a 3 graph-embedded manifold) is a 4-regular properly edge-colored graph such that 
\begin{equation}
v+t=b 
\label{bubbspheres}
\end{equation}
A 4-regular properly edge-colored graph for which (\ref{bubbspheres}) does not apply is called \textit{3-gepm} (a 3 graph-embedded pseudo-manifold).\\\ \\
\textbf{Lemma 2} A necessary and sufficient condition for the graph $(\Gamma,\gamma)_4$ to represent a manifold, is to meet the relation between its 2- and 3- residues (read as it 2- and 3- colored bubbles) and the number of vertices 
(read as the perturbative order) $v+t=b$.\\\ \\
This Lemma clarifies the reason why 3-gems have to satisfy the relation (\ref{bubbspheres}). Let now introduce few definitions which will turn useful later\cite{surveyger}:\\\ \\
\textbf{Definition} A \textit{triball} is a connected, cubic, 3-edge-colored graph $\Gamma_3 \subset \Gamma$ such that its Euler characteristic is the one of the 2-sphere.\\\ \\
Thus we have the relation between its 2-residues $b_{\Gamma_3}$ and the vertices: $ 2 b_{\Gamma_3} - v = 4$. An important fact is the following:\\\ \\
\textbf{Lemma 3} A graph $(\Gamma,\gamma)_4$ is a 3-gem iff each of its 3-residue is a triball.\\\ \\
Thus, the condition that graphs have to satisfy in order to be 3-gems is a condition on the topology of its 3-residues. We now discuss \textit{crystallizations} of 3-gems.
\begin{figure}[htb]
\center
\includegraphics[scale=0.6]{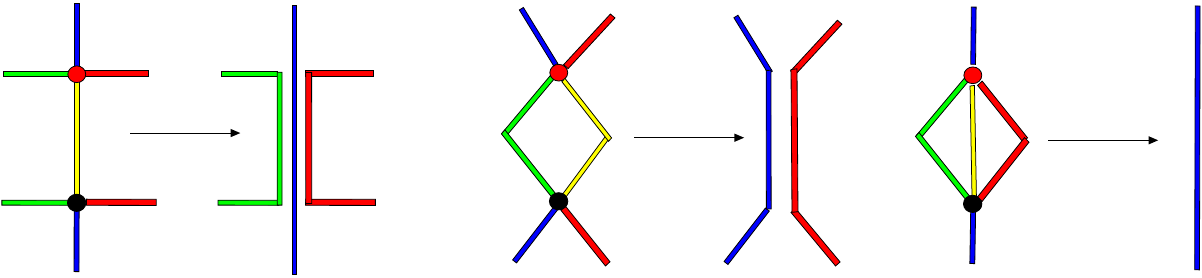}
\caption{Fusion moves on a 4-regular 4-edge colored graph of 1-, 2- and 3- dipoles respectively.}
\label{fusionm} 
\end{figure}
Let first introduce the \textit{fusion} process. Let be $\mathscr B_{ijk}$ and $\mathscr {B'}_{ijk}$ two different 3-residues separated by a unique color which, by construction, is different from the color $i$, $j$, $k$. We call $1-dipole$ this
edge connecting the two 3-residues. The generalization to $k-dipoles$ which connect $(n-k)$-residues is obvious. We call \textit{fusion} the process of contraction of two vertices  
through the first two combinatorial moves depicted in Fig. \ref{fusionm}. Each cancellation of a 1-dipole has the effect of decreasing by one the number of $i$-residues, where i is the color of the edge which defines the 1-dipole, not changing 
the number of $j$-residues,
for $j\neq i$. Thus by a succession of 1-dipole cancellation we obtain a 3-gem with 4 triballs. Such a 3-gem is said to be \textit{contracted} and is called a \textit{crystallization} for the associated 3-manifold.
It is a fact that \textit{any} closed 3-manifolds has a crystallization, and two closed 3-manifolds are related by a homeomorphism if and only if they are related by creation or contraction of 1- and 2- dipoles
with the fusion rules; in this case, the two 3-manifolds are said to be \textit{equivalent} or homeomorphic. Thus it is easy to understand that the fusion rules are the combinatorial equivalent of homeomorphisms.
Let now discuss crystallization for generic colored $(n+1)$-graphs. The following results hold:\\\ \\
\textbf{Theorem 1} For every PL $n$-manifold $\mathscr M$ there exist a crystallization.\\\ \\
\textbf{Theorem 2} Two $n$-graphs $|\Gamma_1|$ and $|\Gamma_2|$ are crystallizations of the same manifold $\mathscr M$ if one 
is converted into the other by:\\
\textit{a}) Adding or removing a non-degenerate m-dipole with $n-1>m>1$;\\
\textit{b}) Adding a $1$-dipole and deleting another $1$-dipole.\\\ \\
A general theorem on the orientability of $n$-graphs holds:\\\ \\
\textbf{Theorem 3 (Orientability)} Let $(\Gamma,\gamma)_{n+1}$ be any crystallization of an $n$-manifold $\mathscr M$. Then $\mathscr M$ is orientable \textit{iff} $\Gamma$ is bipartite.\\\ \\
These theorems are fundamental in order to have a clear geometrical understanding of graphs generated by a colored group field theory, and will be used in the next section, in which the main result of the paper
is presented.
\section{Orientability in cGFT}
In this section we prove a Lemma on the orientability of PL manifolds associated to graphs generated by the colored Boulatov model. Orientability of a manifold is a requirement if we want to construct a spin bundle. In 4-dimensions, for instance,
the requirement to have a global spin bundle is to have a vanishing first and second Stiefel-Whitney class. While the second can be neglected by constructing local spin bundle and then gluing the charts, the vanishing of the first is a strict requirement and is equivalent to ask
the orientability of the manifold\cite{nakahara}. Another important fact is that orientability restricts enormously the class of 3-manifolds which could be generated. As an example,
in 2-dimensions the most general decomposition is given by connected sum of spheres, torii and projective planes. Orientability excludes the connected sum of projective planes, which allows the expansion in the ordinary genus we are used to.\\\ \\
\textbf{Lemma (3-dimensions)} Let $\Gamma$ be a connected vacuum finite graph generated by the colored Boulatov model. Let $\mathscr B_{ijk}$ and $\mathscr B_{ij}$ be the set of 3- and 2- bubbles of $\Gamma$ respectively. Then the pseudo-manifolds
associated to the $\Gamma$ is an \textit{oriented} pseudo-manifold. Moreover, $|\Gamma|$ represents a closed and orientable 3-manifold iff 
\begin{equation}
V+ Card\{\mathscr B_{ijk}\}=Card\{\mathscr B_{ij}\}
\label{beq} 
\end{equation}

\textit{Proof}. This lemma follows directly from the properties of graphs generated by the colored Boulatov model and its interpretation, which is the same of the simplicial construction of \textit{3-gepms}. 
By Lemma 1 the graph generated is a manifold if and only if the condition (\ref{beq})
is met. Since the graph is finite, the manifold is also closed. Thus what we have to show is that they are orientable. By the theorem on the orientability the 3-gem represents an orientable manifold if and only if the crystallization graph is bipartite. First we note that the graphs generated by colored group field theory are bipartite. 
Let $A$ and $B$ be the set of clockwise and anti-clockwise vertices of $\Gamma$ respectively. Since by construction a clockwise vertex
has to be contracted with an anticlockwise, then all the edges are between the set $A$ and the set $B$ and none is within the sets, thus the graph is bipartite. 
Now we have to show that its contraction is still bipartite. However, this fact is trivial because any of the moves in Fig. \ref{fusionm} keeps the bipartiteness of the original graph, thus in particular
the fusion of a $1-dipole$. Moreover, since the graph is finite, the crystallization is reached in a finite number of moves.\\\ \\
The orientability part of this Lemma can be generalized to higher dimensions. The construction given in the third section of this note ensures that to each $n$-dimensional pseudo-complex there is at least a colored $(n+1)$-graph which is homehomorphic to it.
It is then easy to see why colored group field theories generates only orientable pseudo-manifolds in any number of dimensions; we state it as a Lemma, even if it clearly follows from the construction given in \cite{Pezzana}
of $n$-edge-colored graphs in any number of dimensions, while orientability comes from a generalization to m-dipoles (as in Theorem 2) of the previous proof and the fact that there are two types of vertices:
This means that, at any finite order, the connected vacuum graphs generated by the partition function of a colored group field theory are associated with closed and orientable PL pseudo-manifolds.\\

\section{Conclusions}
In this short paper we have used results in the field of 3-gems to prove that all the graphs generated by the colored Boulatov model are related to orientable pseudo-manifolds.
In order to prove it we used new tools which could turn to be very useful in the context of group field theory, more specifically in the \textit{colored} version of it.
In fact, color is a fundamental ingredient in all we said. It should be said that what proved here is not an unexpected result\cite{talks}. The fact that an orientation for the faces can be chosen with ease was a hint
of what proved here. Indeed, as far as the author is concerned, this is the first rigorous proof appeared so far. 
We should stress that orientability is a fundamental requirement for ``reasonable'' manifolds.
\begin{center}
\textbf{Aknowledgements} 
\end{center}

We would like to thank Razvan Gurau for several discussions on the topic of group field theory. Also, we would like to thank Daniele Oriti for advices on the presentation of this
result and Lorenzo Sindoni for reading carefully the manuscript.
Research at Perimeter Institute for Theoretical Physics is supported in part by the Government of Canada through NSERC and
by the Province of Ontario through MRI. 
\begin{center}

\end{center}

%

\end{document}